\documentclass[twocolumn,showpacs,preprintnumbers,amsmath,amssymb,superscriptaddress]{revtex4}

\usepackage{graphicx}% Include figure files 
\usepackage{dcolumn}% Align table columns on decimal point 
\usepackage{bm}% bold math
\usepackage{hyperref} 
\usepackage{latexsym}

\begin{document}

\title{Extremum statistics in scale-free network models}

\author{Andr\'{e} Auto \surname{Moreira}} 
\email{auto@fisica.ufc.br}

\affiliation{Center for Polymer Studies and Dept. of Physics, Boston
		University, Boston, MA 02215}

\affiliation{Departamento de F\'{\i}sica, Universidade Federal do 
		Cear\'a, 60451-970 Fortaleza, Cear\'a, Brazil}

\author{Jos\'{e} S. \surname{Andrade} Jr.}
\email{soares@fisica.ufc.br}
\affiliation{Departamento de F\'{\i}sica, Universidade Federal do
		Cear\'a, 60451-970 Fortaleza, Cear\'a, Brazil}  

\author{Lu\'{\i}s A. Nunes \surname{Amaral}} 
\email{amaral@buphy.bu.edu}
\homepage{http://polymer.bu.edu/~amaral} 
\affiliation{Center for Polymer Studies and Dept. of Physics, Boston
		University, Boston, MA 02215}

\begin{abstract}

We investigate the statistics of the most connected nodes in scale-free
networks. For a scale-free network model with homogeneous nodes, we show by
means of extensive simulations that the exponential truncation---due to the
finite size of the network---of the degree distribution governs the scaling
of the extreme values.  We also find that the distribution of maxima obeys
scaling and follows the Gumbel statistics.  For a scale-free network model
with heterogeneous nodes, we show that scaling no longer holds and that the
truncation of the degree distribution no longer controls the maximum
distribution.  Moreover, we find that neither the Gumbel nor the Frechet
statistics describe the data.

\end{abstract}


\maketitle

%%%%%%%%%%%%%%%%%%%%%%%%%%%%%%%%%%%%%%%%%%%%%%%%%%%%%%%%%%%%%%
%%%%%%%%%%%%%% Introduction

The statistics of extrema is a classical subject of great interest in
mathematics and physics \cite{extrem}. In physics, extreme events have been
studied in a number of fields, including self-organized fluctuations and
critical phenomena \cite{bramwell98}, material fracture \cite{sahimi93},
disordered systems at low temperatures \cite{bouch}, and turbulence
\cite{outiers}. Knowledge of extreme event statistics is also of fundamental
importance to predict and estimate risk in a variety of natural and man-made
phenomena such as earthquakes, changes in climate conditions, floods
\cite{clime}, and large movements in financial markets \cite{market}.  A new
field where extreme statistics is of interest is complex networks
\cite{new}. For one particular class of complex networks
\cite{class}---scale-free networks \cite{AB1999,rev}---it is well known that
the most connected nodes strongly influence the dynamics of the system,
playing a fundamental role in many different phenomena such as Internet
response to attacks \cite{AJB00}, spreading of epidemics \cite{epdemy}, or
propagation of email virus \cite{virus}. Surprisingly, so far there has been
no attempt to characterize the distribution of extreme connectivities in
scale-free networks.

An important result in extreme statistics is that the distributions of maxima
for independent identically-distributed ({\it iid\/}) random variables fall
onto a small number of universality classes \cite{extrem}.  Let $C \equiv
\{k_1,k_2,..., k_S\}$ be a set of {\it iid\/} variables drawn with
probability density function $p(k)$. The distribution $\rho(K)$ of the
maximum $K$ in the set $C$ is dictated by the asymptotic behavior of the tail
of $p(k)$ \cite{extrem}.  Specifically, $\rho(K)$ converges to the Gumbel
distribution,
\begin{equation} 
 \rho(K)=a \exp(-u-e^{-u}), 
 \label{gumbel}
\end{equation} 
where $u=a(K-b)$, when $p(k)$ decays faster than a power law; and to the
Frechet distribution,
\begin{equation}
 \rho(K)=\alpha K^{-(\alpha+1)}\exp(-K^{-\alpha}),  
 \label{frechet}
\end{equation} 
when $p(k)$ decays as $k^{-(\alpha+1)}$ \cite{weibul,bou}.

Unlike the case of {\it iid\/} random variables, little is known when
correlations are present among the variables $k_i$ \cite{bouch,dean01}. Even
though the universality classes of uncorrelated and correlated variables may
not necessarily be the same, correlated systems have been generally studied
under the framework of {\it iid\/} extreme statistics \cite{bazant}. In the
case of scale-free networks, which we consider in this Letter, correlations
are present in the degrees $k_i$ (i.e. the number of links) of the nodes
\cite{constr}.

To investigate the extreme statistics in scale-free networks, we consider
here the fitness model of Ref~\cite{bianc2001}.  This model is a
generalization of the scale-free model of Ref.~\cite{AB1999}, in which nodes
have heterogeneous fitnesses $\{\eta_i;~ i = 1,...,S\}$. The fitness $\eta_i$
models an inherent quality of the node $i$ that ``weighs'' its attractiveness
to new links.  In this model, the network starts with $s_0$ nodes, each with
$s_0-1$ links. At time $t$, a new node is added to the network and
establishes $s_0-1$ new links. A new link is established with a node $i$,
from the set of the $t-1+s_0$ existing nodes, with a probability proportional
to the node degree $k_i$ and fitness $\eta_i$
\begin{equation} 
 \Pi_i=\frac{k_i \eta_i}{\sum{k_j \eta_j} } \,.
\end{equation}
This mechanism, typically denoted ``preferential attachment,'' drives the
network to a degree distribution that decays in the tail as a power law
\cite{AB1999}. In the homogeneous case, $\eta_i=1$ for all $i$, one recovers
the original model of Ref.~\cite{AB1999}, and generates a network with
cumulative degree distribution that decays as $P(k) \sim k^{-2}$
\cite{AB1999}. In the other case we study here, the fitnesses $\eta_i$ are
drawn from a uniform distribution $\eta_i~ \epsilon~ [0,1]$. This case
generates a network with a cumulative degree distribution of the form
$P(k)\sim k^{-\alpha}/\log(k)$, with $\alpha=1.225$ \cite{bianc2001}.

For the case of nodes with homogeneous fitness---i.e., all $\eta_i$ are
equal---we find that the distribution of maxima obeys Gumbel statistics. This
is a surprising result for two reasons: (i) the degrees $k_i$ are not {\it
iid\/} variables and are not equally distributed---hence, there is not an
{\it a priori\/} justification to expect that one of the two universality
classes (1) or (2) will hold---and (ii) the distribution $p(k)$ decays as a
power law---hence, one would more likely expect to find Frechet statistics.
For the case of nodes with heterogeneous fitness, we find (i) absence of
scaling, i.e., the shape of the distribution changes with the network size,
and (ii) that the distribution of maximum, for finite network sizes, is not
consistent with either of the universality classes represented by
Eqs.~(\ref{gumbel}) and (\ref{frechet}).  However, our results are consistent
with the possibility that the distribution of maxima may converge to the
Frechet distribution in the thermodynamic limit.  We show that the absence of
scaling for the heterogeneous fitness case is due to the progressive entry in
the system of nodes with larger fitness that eventually become the new
maxima.

%
%%%%%%%%%%%%%%%%%%%%%%%%%%%%%%%%%%%%%%%%%%%%%%%%%%%%%%%% FIGURE 1
\begin{figure}[tb] 
 \vspace*{0.cm}
 \includegraphics*[width=6.cm]{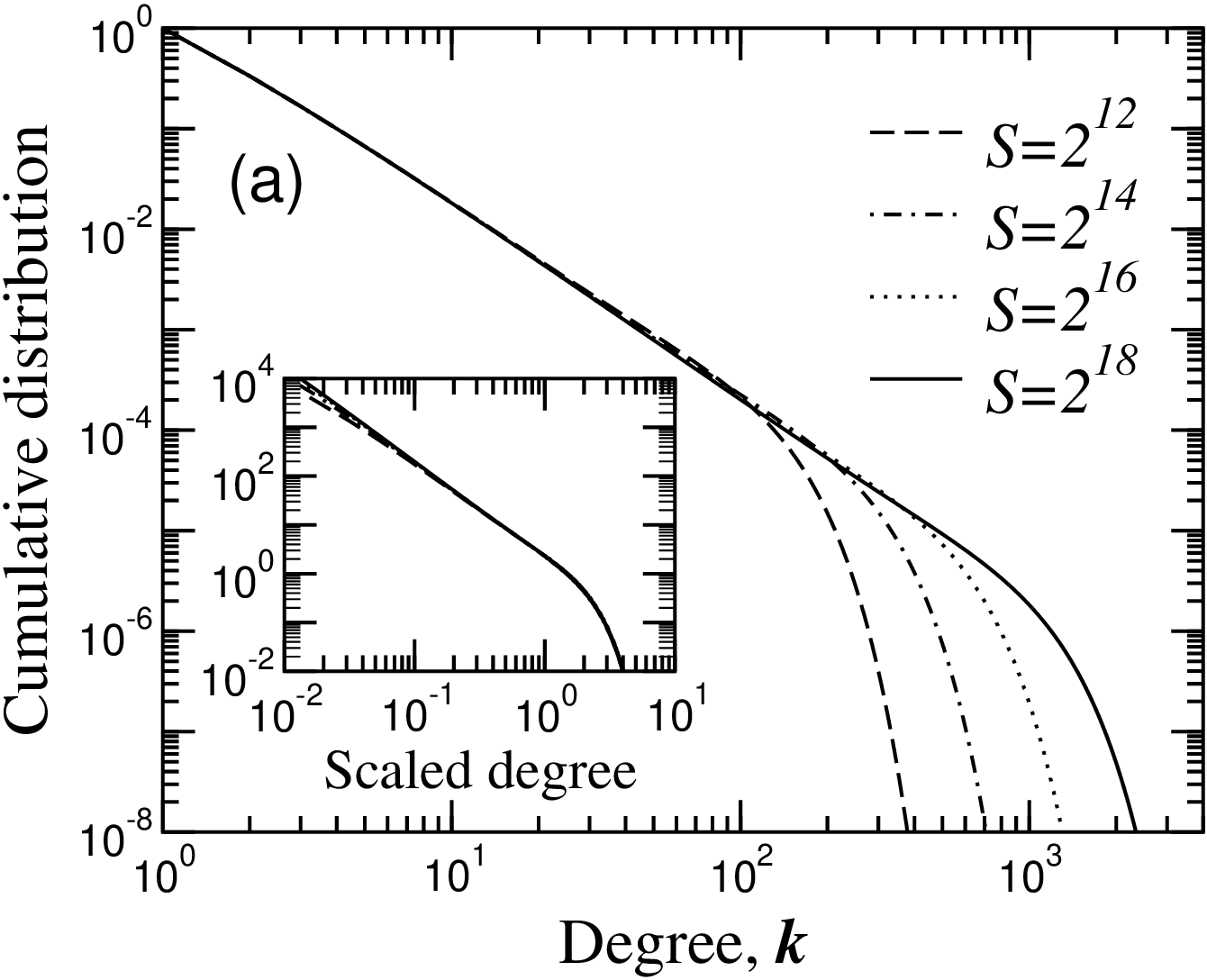}

 \vspace*{0.3cm} 
 \includegraphics*[width=6.1cm]{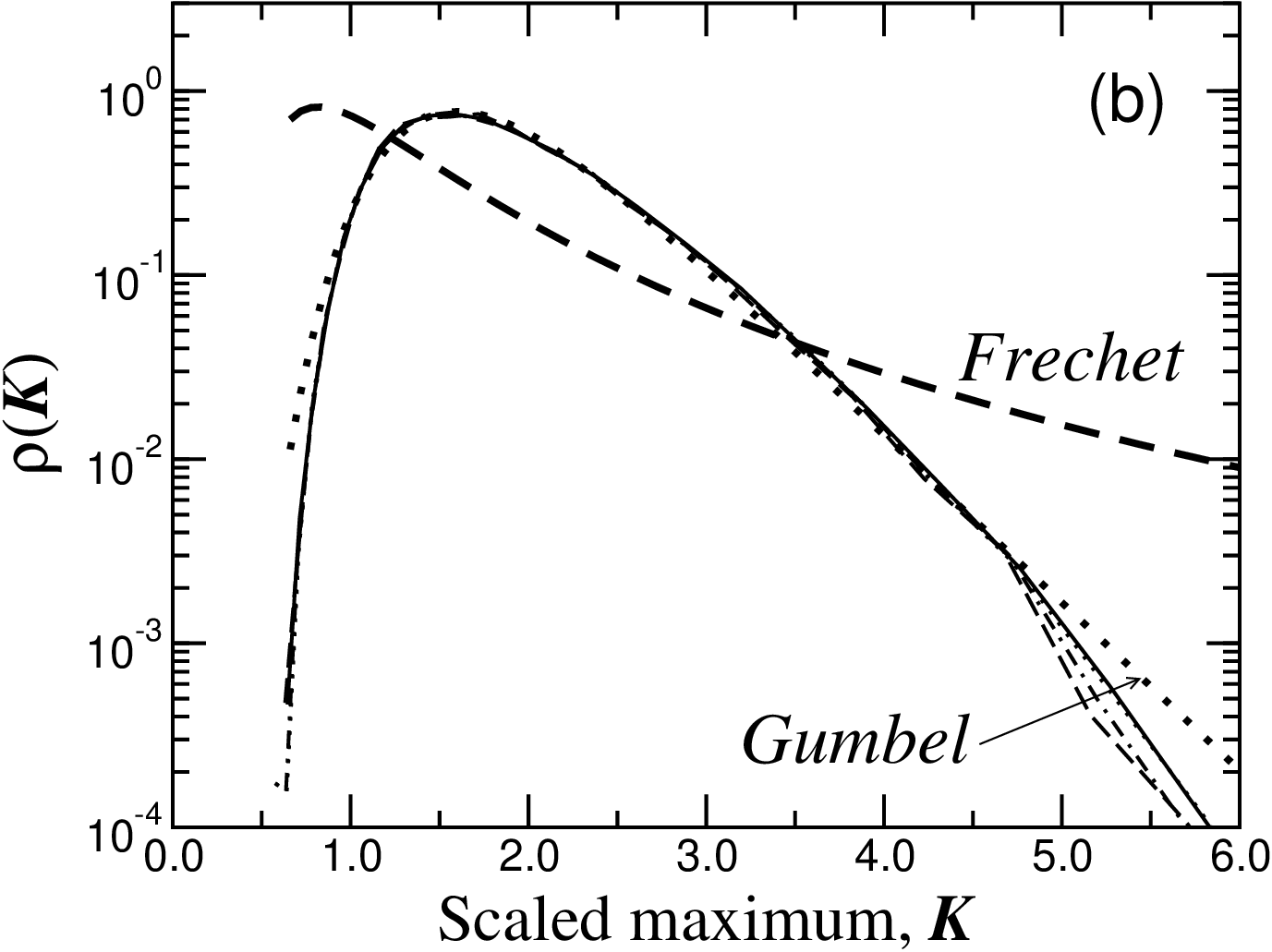}

 \vspace*{-0.cm} 
 \caption{(a) Cumulative degree distribution for the case $\eta_i=1$,
	corresponding to the scale-free model of Ref.~\cite{AB1999}.  The
	cumulative distribution decays as a power-law with exponent
	$\alpha=2$, followed by an exponential truncation. The inset shows
	the data collapse obtained by the rescaling $S^{\alpha \theta}
	P\propto\left(k / {S^{\theta}} \right)^{-\alpha}$, where $\theta= 1/
	\alpha = 0.50 \pm 0.03$ is the exponent controlling the onset of the
	exponential truncation \cite{dorog2001,krap2000}.
	(b) Distribution of the maximum degree.  We generate $10^5$ network
	realizations for each size $S$ and confirm that the results are
	indistinguishable from those obtained for $10^4$ network
	realizations. In all simulations the initial size is $s_0=2$.  We
	plot the distribution functions of the scaled maximum $K^{\prime}
	\equiv S^{-\gamma}K$, where $\gamma=0.50\pm 0.03$.  As expected, the
	distribution of maxima displays the same scaling as the truncation of
	the degree distribution i.e., $\gamma \approx \theta$.  Also shown is
	the fittings of the data to the Gumbel with $u=-2.1(K^{\prime}-1.6)$,
	and Frechet distributions with $\alpha=2$.  The maximum statistics
	agrees well with the Gumbel distribution for $K^{\prime}<5$.  }
 \label{nof} 
\end{figure}
%%%%%%%%%%%%%%%%%%%%%%%%%%%%%%%%%%%%%%%%%%%%%%%%%%%%%%%%%%%%%%%%%%%
%

%%%%%%%%%%%%%%%%%%%%%%%%%%%%%%%%%%%%%%%%%%%%%%%%%%%%%%%%%%%%%%%%%%%%
%%%%%%%%%%% RESULTS

First we consider the case $\eta_i=1$ for all $i$. The distribution of the
maximum degree $K$ is non trivial because (i) the degrees $k_i$ display a
constraint on the total number of links, and (ii) the variables $k_i$ are not
identically distributed. Indeed, recent studies have shown that each node has
a different probability distribution for its degree $p_i(k_i)$ obeying an
exponential form with a characteristic scale that depends on the square root
of the node index $i$ \cite{dorog2000,krap2000}.

%
%%%%%%%%%%%%%%%%%%%%%%%%%%%%%%%%%%%%%%%%%%%%%%%%%%%%%%%%%% FIGURE 2
\begin{figure}[tb] 
 \vspace*{0.cm}
 \includegraphics*[width=6.cm]{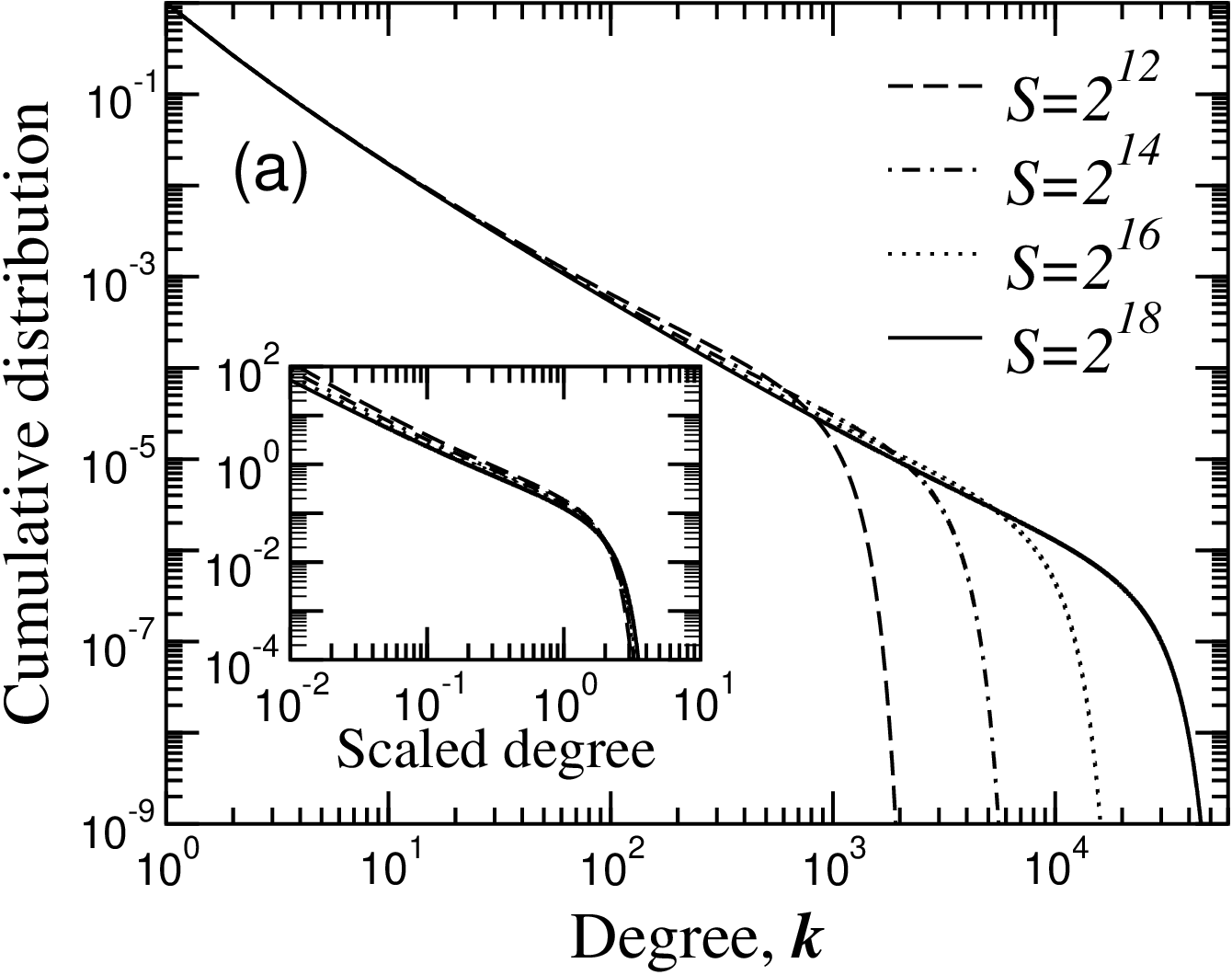}

 \vspace*{0.3cm}
 \includegraphics*[width=6.1cm]{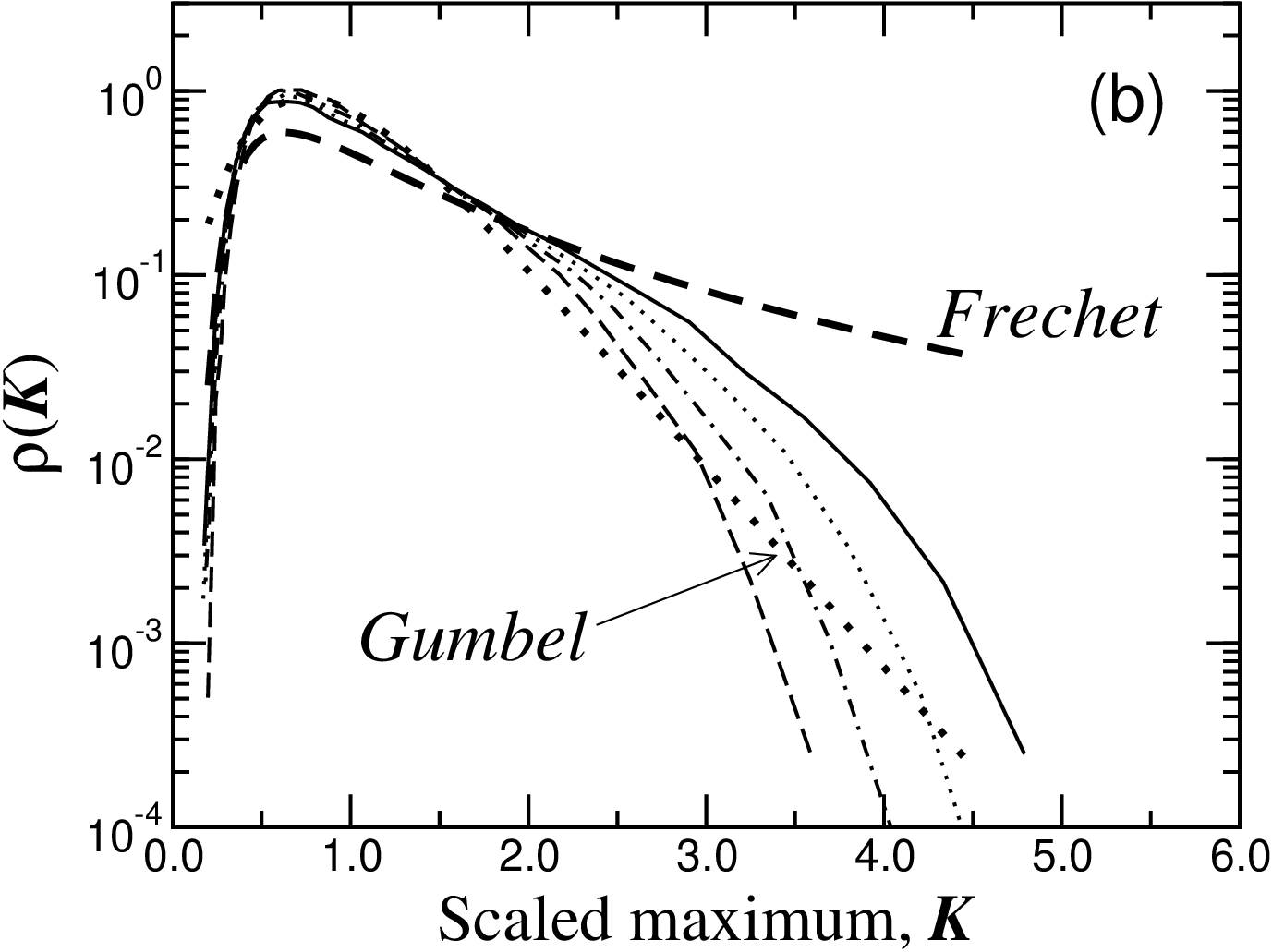}

 \vspace*{-0.cm} 
 \caption{(a) Cumulative degree distribution for the model of
  	Ref.~\protect\cite{bianc2001} with uniform fitness distribution.  In
  	this case, the distributions decay with an exponent $\alpha=1.255$
  	with logarithmic corrections \protect\cite{bianc2001}. Note that this
  	result is different from the results of Fig.~2a of
  	\protect\cite{bianc2001} which shows a plateau instead of an
  	exponential truncation.  The inset shows the curves rescaled in the
  	same way as in Fig.~\protect\ref{nof}(a), but with an exponent
  	$\theta=0.76 \pm 0.05$.
	(b) Data collapse for the maximum degree distribution, obtained from
	$10^5$ network realizations for each size. We use the rescaling
	relation $K^{\prime} \equiv S^{-\gamma}K$, with $\gamma=0.7 \pm 0.1 $
	to collapse the data.  The thin dotted and dashed lines are the
	fitting of the data for the Gumbel with $u=-2.5(K^{\prime}-0.75)$,
	and Frechet with $\alpha=1.255$. For this case the curves do not
	collapse well.  On the contrary, the distributions become broader
	as the network grows, appearing to converge to the Frechet
	distribution as $S$ increases.
%The figure clearly shows that
%for the fitness model, the connectivity and maximum distributions
%follow different scaling laws.
%%%%%%%%%%%%%%%%%%%%%%%%%%%%%%%%%%%%%%%%%%%%%%%%%%%%%%%%%%%%%%%%%%%%%
% Do you mean that the data collapse is also not perfect?
%%%%%%%%%%%%%%%%%%%%%%%%%%%%%%%%%%%%%%%%%%%%%%%%%%%%%%%%%%%%%%%%%%%%%
}  
 \label{fit}
\end{figure} 
%%%%%%%%%%%%%%%%%%%%%%%%%%%%%%%%%%%%%%%%%%%%%%%%%%%%%%%%%%%%%%%%%%
%

%These facts raise an important question: One expects the maximum statistics
%to be governed by the tail of the degree distribution.  In this case,
%however, it is not clear which behavior will dictate the maximum statistics,
%the one-node degree distribution $p_i(k_i)$ or the global degree distribution
%$p(k)$? The exponential decay of $p_i(k_i)$ suggests that the distribution of
%maxima converges to the Gumbel, Eq.~(\ref{gumbel}), while the power law decay
%of $p(k)$ suggests that the distribution of maxima should take the broader
%form of the Frechet distribution, Eq.~(\ref{frechet}).

Figure \ref{nof}(a) shows the cumulative degree distribution, $P(k) =
\sum_{k'>k} p(k')$, for different network sizes $S$ for the homogeneous case,
$\eta_i=1$.  The curves where obtained by numerically iterating the rate
equation proposed in Ref~\cite{krap2000}.  As expected, the curves display a
power law decay, $P(k) \sim k^{-\alpha}$, with $\alpha=2$, followed by an
exponential truncation. The inset shows the data collapse obtained by
rescaling all curves according to
\begin{equation} 
 S^{\alpha \theta} P \propto
 \left(\frac{k}{S^{\theta}}\right)^{-\alpha} 
 \label{scaling}
\end{equation} 
where $\theta = 0.50 \pm 0.03 = 1/\alpha = 1/2$ is the exponent that governs
the onset of the exponential truncation \cite{dorog2001}. Figure \ref{nof}(b)
shows the distribution of maximum degree rescaled as $K^{\prime} \equiv
S^{-\gamma}K$, with $\gamma=0.50 \pm 0.03$.  Also shown in Fig.~\ref{nof}(b)
are the best fittings of the data to the Frechet and Gumbel distributions. It
is visually apparent that the Gumbel distribution describes the maximum
statistics surprisingly well for $K^{\prime} < 5$ \cite{note33}.

This is a surprising result for two reasons: First, we find Gumbel statistics
even though the degrees $k_i$ are not {\it iid\/} variables and are not
equally distributed.  Hence, our results appear to indicate that in this case
these facts do not affect the maximum statistics.  Second, we find Gumbel
statistics even though $p(k)$ decays as a power law which would lead one to
expect Frechet statistics.  The reason for not finding Frechet statistics is
the fact that the network growth process generates finite size networks.
This contrasts with the procedure when drawing a set of $S$ variables from a
given distribution.  Specifically, if one draws ten samples of size 1000,
this is no different from drawing a single sample of size 10,000.  In
contrast, in the case of a scale-free network, there is a limit for the
maximum degree possible \cite{note_test} that is controlled by the exponent
$\theta$; i.e., $K \sim S^{\theta}$ \cite{dorog2001,krap2000}.  Hence, for
scale-free networks, generating ten networks of size 1000 leads to a
statistically different set of $k_i$'s than generating one network of size
10,000.  In the former case $K \sim 1000^{0.5} \approx 30$, while in the
latter $K \sim 10000^{0.5} \approx 100$.  Because for scale-free networks
there is an exponential truncation due to finite network size of $p(k)$, it
is natural that we find Gumbel statistics of the maxima and that $\theta =
\gamma$.

%
%%%%%%%%%%%%%%%%%%%%%%%%%%%%%%%%%%%%%%%%%%%%%%%%%%%%%%% FIGURE 3
\begin{figure}[t]  
 \vspace*{0.cm}
 \includegraphics[width=6.1cm]{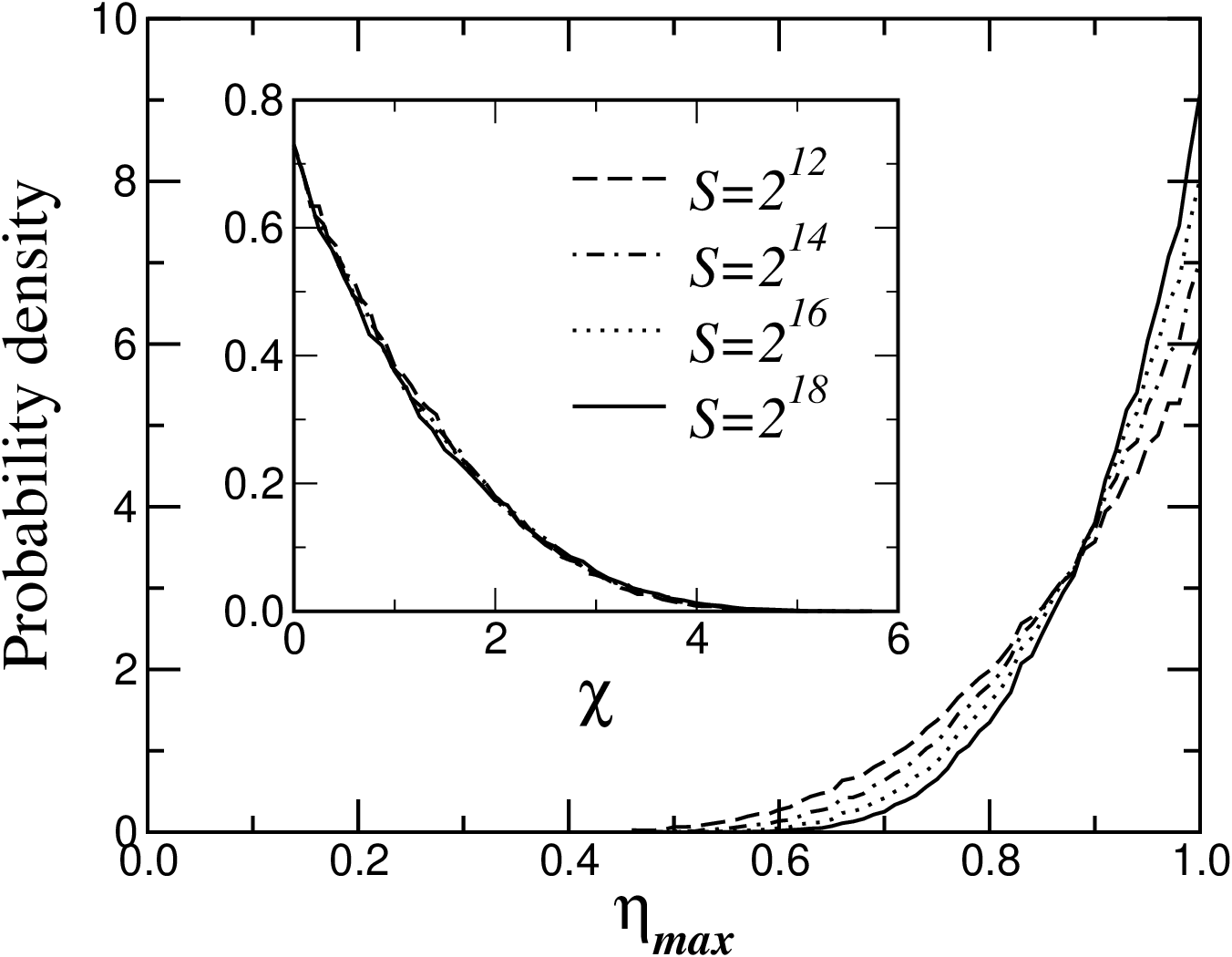}
 \caption{ Distribution of the fitness $\eta_{max}$ of the node with the
	maximum degree. As expected, the node with the largest fitness
	eventually becomes the most connected during the growth of the
	network. The inset shows the collapse of the data performed under the
	transformation $\chi \equiv (1-\eta_{max}) \log S$.  }
 \label{fitmax} 
\end{figure}
%%%%%%%%%%%%%%%%%%%%%%%%%%%%%%%%%%%%%%%%%%%%%%%%%%%%%%%%%%%%%%%%
%

We next consider the case with uniform distribution of fitnesses. As before,
we study the degree distribution and the maximum statistics \cite{how}.  As
shown in Fig~\ref{fit}(a), the cumulative degree distributions follow the
expected scaling and display, as in the previous case, an exponential
truncation that scales as $S^{\theta}$ with $\theta = 0.76 \pm 0.05$.  In
Fig.~\ref{fit}(b) we show the rescaled distributions of maximum degree for
different network sizes $(K^{\prime}=S^{-\gamma}K)$ with $\gamma = 0.7 \pm
0.1$.  Although the data collapse is not perfect, the estimates for the
exponents $\theta$ and $\gamma$ are within statistical uncertainties and in
agreement with the conjectured value $\theta = 1 / \alpha = 1 / 1.255 \approx
0.786$ \cite{dorog2001,krap2000}.

%
%%%%%%%%%%%%%%%%%%%%%%%%%%%%%%%%%%%%%%%%%%%%%%%%%%%%%%%%% FIGURE 4
\begin{figure}[bt] 
 \vspace*{0.cm}
 \includegraphics*[width=6.cm]{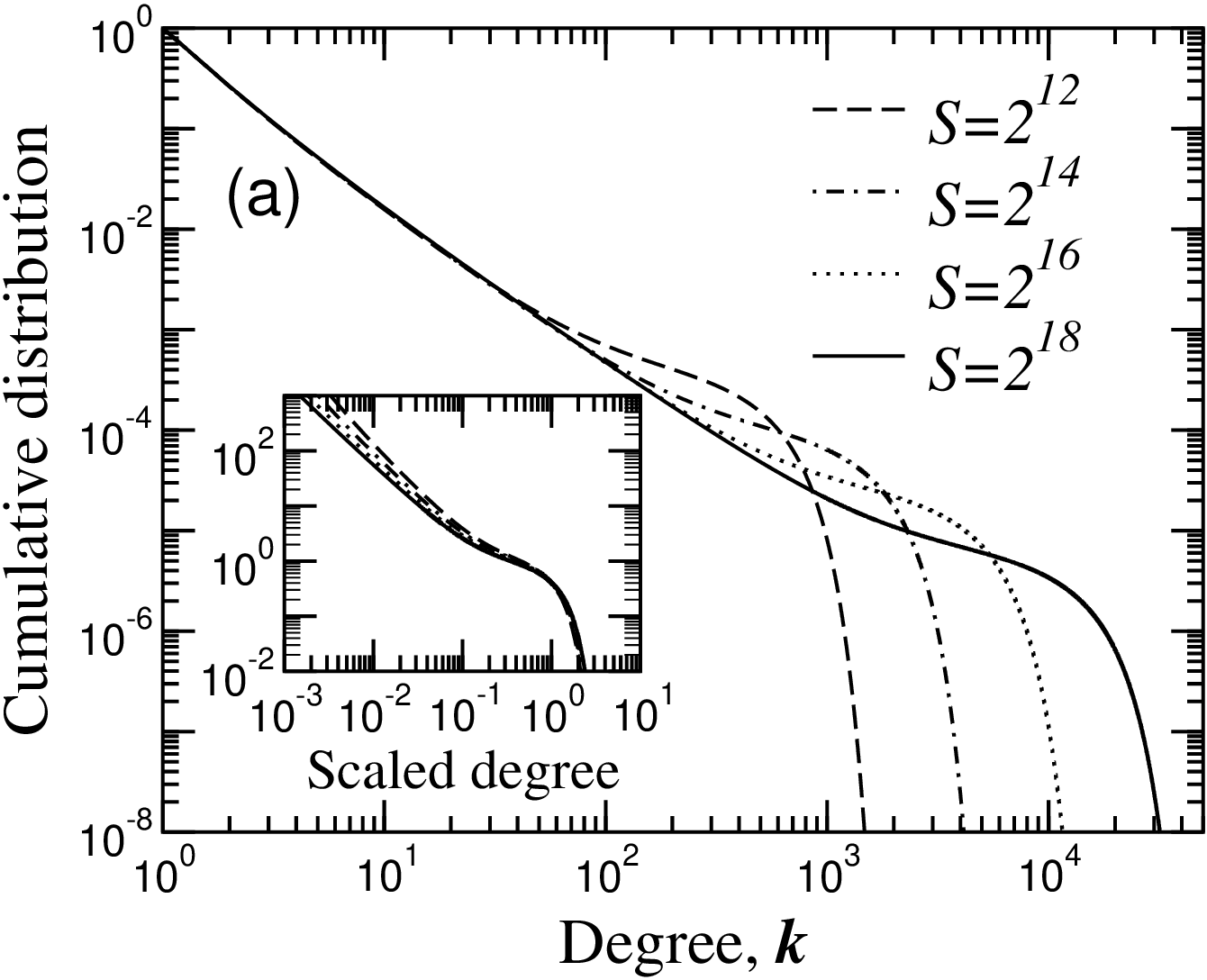}

 \vspace*{0.3cm}
 \includegraphics*[width=6.1cm]{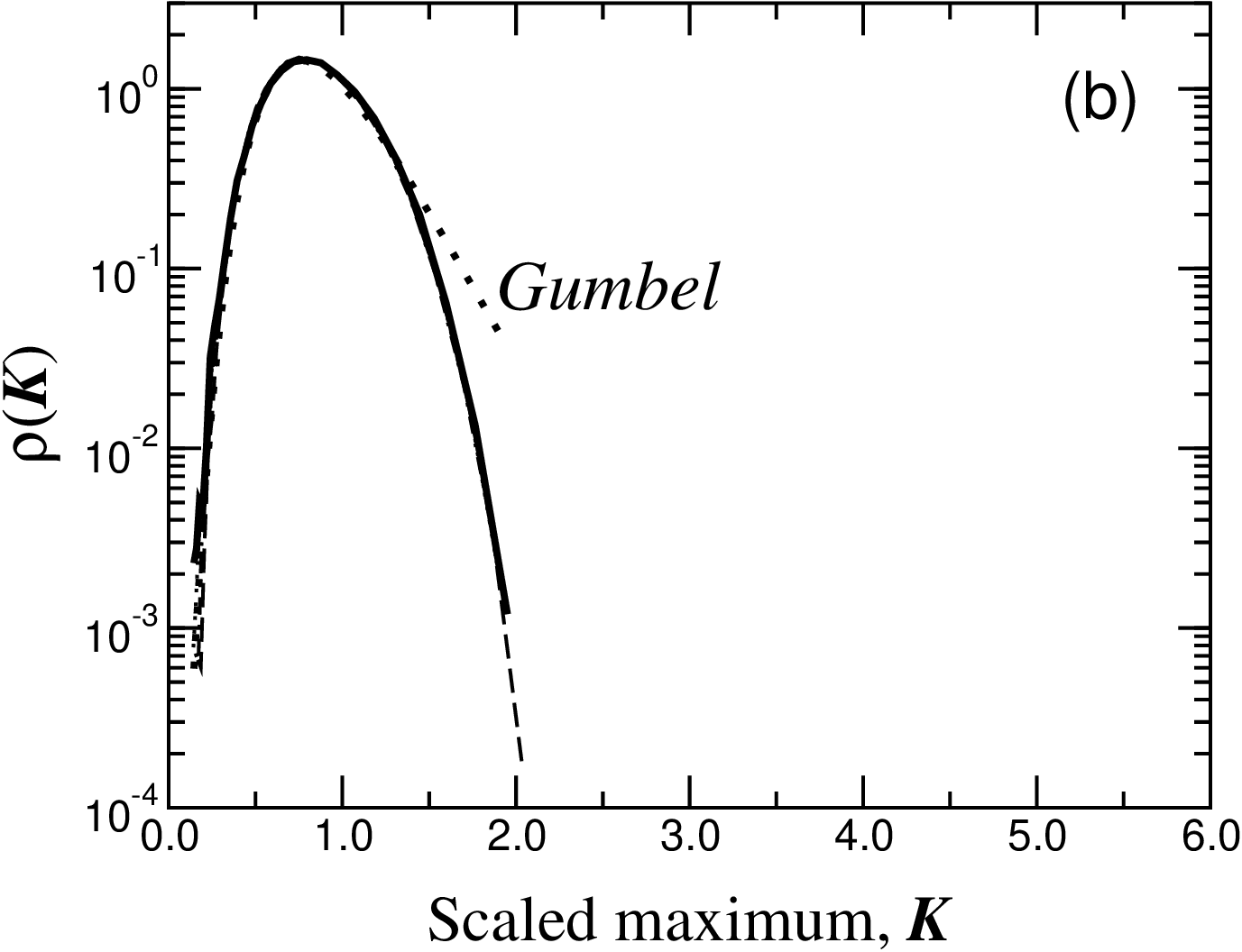}

 \caption{ The results shown in this figure correspond to a test with the
	fitness model where we arbitrarily set the two first nodes present in
	the beginning of the dynamics to have $\eta=1$.  (a) Cumulative
	degree distribution. The data collapse shown in the inset has been
	obtained in the same fashion as Figs.~\protect\ref{nof}(a) and
	~\protect\ref{fit}(a), with $\alpha=1.27 \pm 0.05$ and $\theta=0.78
	\pm 0.04$.
	(b) Distribution of maximum degree averaged over $10^5$ network
	realizations.  
%Data collapse of the distributions of maxima done
%	with uniform fitness distribution.
%In this case the fitness of
%the the two first nodes present in the beginning of the
%dynamics are set to have the maximum possible value of
%fitness. 
	The rescaling relation $K^{\prime} \equiv S^{-\gamma}K$, with
	$\gamma=0.78 \pm 0.04$ has been used.  The scaling exponent is the
	same that we found for the degree distribution in this case. This
	supports the hypothesis that the ``anomalous'' behavior of the
	extreme statistics in the fitness model is due to the introduction of
	a vertex with large fitness in the network which subsequently
	overtakes the position of the highest-degree node.  The thin dotted
	line is the best fit to the data of the Gumbel distribution with
	$u=-4(K^{\prime}-0.76)$.  }
 \label{tes} 
\end{figure}
%%%%%%%%%%%%%%%%%%%%%%%%%%%%%%%%%%%%%%%%%%%%%%%%%%%%%%%%%%%%%%%%%%%%%%
%

In Fig.~\ref{fit}(b) we also show the best fitting to the data of the Gumbel
and Frechet distributions.  The results do not follow any of the two
classical distributions, but the curves appear to converge to the Frechet
distribution as $S$ increases.

In order to understand the effect of fitness in the growth model, we compute
the fitness distribution of the most connected node. Figure~\ref{fitmax}
shows that, as the network grows, nodes with increasing fitness become the
ones with maximum degree. This is to be expected since the growth of the
degree of a node increases over time as a power law with an exponent
proportional to its fitness \cite{bianc2001}.  Indeed, as the network grows,
the distribution converges logarithmically to a delta function at
$\eta=1$. Based on this fact, we may then suppose that the failure to
collapse the data in Fig.~\ref{fit}(b) may be a consequence of the slow
progressive entry into the system of nodes with larger fitnesses which
eventually overcome older nodes that had the largest degree.

In order to test this hypothesis, 
%that the late entry of higher fitness nodes
%is responsible for the absence of scaling of the distribution of maxima in
%the model of Ref.~\cite{bianc2001}
we consider an additional case of heterogeneous nodes where the two first
sites of the growing network ($i=1$ and $2$) are set to have fitness one,
while all other nodes have fitnesses drawn from a uniform distribution.  This
case implies that one of the two first sites will become the node with
maximum degree and that the distribution of $\eta_{max}$ is a delta function
at one.

We calculate the cumulative degree distribution for this case and find that
the distribution displays a power law decay followed by a short plateau just
before the exponential truncation; cf. Fig~\ref{tes}(a).  We find that the
distributions obtained for the different network sizes can be collapsed
according to Eq.~(\ref{scaling}) with the exponents $\alpha = 1.27 \pm 0.05$
and $\theta \approx 0.78 \pm 0.04$ \cite{note55}.  We also calculate the
distribution of maxima and find that the distributions for different $S$ can
be collapsed upon the rescaling $K^{\prime}=S^{-\gamma}K$ with $\gamma
\approx 0.78$. The distributions are consistent with the Gumbel statistics in
the region around the most probable maxima.  Moreover, our estimate of
$\gamma$ is in agreement with the value we obtain for the truncation of the
power law regime of the cumulative degree distribution, $\theta \approx 0.78
\pm 0.04$. These results support our conjecture that the absence of good
scaling in the distribution of maxima for the scale-free model of
Ref.~\cite{bianc2001} is due to the progressive entry of nodes with larger
fitness.

The major finding of this study is that the distribution of maxima for
scale-free models has non trivial properties.  For the case of homogeneous
nodes---i.e., nodes with identical fitness---we find that the distribution of
maxima follows Gumbel statistics with parameters related to the exponent
$\alpha$ characterizing the degree distribution.  We explain this finding by
the exponential truncation of $p(k)$ due to finite network size. In contrast,
for scale-free models with heterogeneous nodes having fitnesses drawn from an
uniform distribution, we find no scaling of the distribution of maxima.  We
explain this lack of scaling in terms of the progressive entry of nodes with
larger fitness which over time will establish more links than nodes with
lower fitness that entered the system earlier.  Surprisingly, our results for
this case of heterogeneous nodes are not inconsistent with the possibility
that the asymptotic distribution of maxima follows the Frechet statistics
even though $p(k)$ is exponentially-truncated due to finite network size.

%%%%%%%%%%%%%%%%%%%%%%%%%%%%%%%%%%%%%%%%%%%%%%%%%%%%%%%%%%%%%
%\acknowledgments 

We thank S. Mossa, and H. E. Stanley for stimulating discussions and helpful
suggestions. A.A.M. and J.S.A. thank the Brazilian Agencies CNPq and FUNCAP
for support. L.A.N.A. thanks NIH/NCRR (P41 RR13622) and NSF for support.

%%%%%%%%%%%%%%%%%%%%%%%%%%%%%%%%%%%%%%%%%%%%%%%%%%%%%%%%%%%%%%%%%%%%%%%%
%%%%%%%%%%%%%%%%%%%%%%%%%%%%%%%%%% REFERENCES


\begin{thebibliography}{99}
\bibitem{weibul} A third case, known as Weibull distribution,
occurs when the distribution follows a positive power law.
In this case the variables have to be bounded \cite{extrem}.
\bibitem{bou} In a recent study, these universality classes 
have been confirmed by means of the replica symmetry
breaking method
\cite{bouch}.
\bibitem{constr} The total number of links is constrained to the
size of the network $\sum{k_i}=2S$.
\bibitem{note33} Our results are consistent with the possibility that, as $S$
increases, there is a very slow convergence to the Gumbel distribution even
for $K'>5$.
\bibitem{note_test} To test this fact, we consider the maximum statistics of
a small sub-set of the nodes for a network of a given size $S=2^{18}$ and as
expected find Frechet statistics for the maximum degree of the sub-set.
\bibitem{how} In order to calculate the cumulative degree distribution in
this model, we discretize the distribution of fitness in the range $[0,1]$ to
$n=100$ equally spaced possible values. We then iterate consistently the rate
equation through this sequence and sum over the fitness values.  We obtain
identical results both for a more fine discretization ($n=1000$) and also for
Monte-Carlo simulations.
\end{thebibliography}
\end{document}